 \documentclass[aps,prd,superscriptaddress,showpacs,twocolumn,preprintnumbers]{revtex4}
\usepackage{epsfig,dcolumn}
\usepackage{graphicx}
\usepackage{color}

\usepackage{amsmath}
\usepackage{amsfonts}
\usepackage{amssymb}
\usepackage{graphicx}

\newcommand{\be}{\begin{equation}}
\newcommand{\ee}{\end{equation}}

\usepackage{bm} 

\begin{document}

\title{Application of the Veneziano Model in Charmonium Dalitz Plot Analysis} 

\author{Adam P. Szczepaniak} 
 \affiliation{ 
  Department of Physics, Indiana University, Bloomington, IN 47405, USA } 
\affiliation{ Theory Center, Thomas Jefferson National Accelerator Facility, \\
12000 Jefferson Avenue, Newport News, Virginia 23606, USA} 
  \affiliation{ 
 Center for Exploration of Energy and Matter, Indiana University, Bloomington, IN 47403, USA}

\author{ M.R. Pennington}
\affiliation{
 Theory Center, Thomas Jefferson National Accelerator Facility, \\
12000 Jefferson Avenue, Newport News, Virginia 23606, USA}

\date{\today}

\begin{abstract}
We adapt the Veneziano model to the analysis of vector charmonium decays. 
 Starting from a set of covariant Veneziano  terms 
we show how to construct partial waves amplitudes that receive contributions from selected Regge trajectories. The amplitudes, nevertheless retain the proper asymptotic limit. This arises from duality between directly produced resonances and cross-channel Reggeon and in practical applications helps remove uncertainties  in the parametrization of backgrounds. 
\end{abstract}

\maketitle

\section{Introduction}
\label{intro} 

We consider the generalized  Veneziano amplitude~\cite{V}  and its  application in analyses of decays of  heavy quarkonia.
Specifically we focus on the decays of vector charmonia, {\it e.g.} $J/\psi$ and $\psi'$ to three pions.
 
 Decays of  charmonia have been investigated by  MARKII, CLEO, BaBar, BES and more recently 
   by BESIII~\cite{1,2,3,4,5,6,7,8}.  One of the original motivations  was to verify perturbative QCD~\cite{QCD}. The QCD calculations are  based on the assumption 
    that the initial  quarkonium wave function and the wave functions of the light hadrons in the final state factorize. The latter is supposed to be produced through a calculable short distance process that follows annihilation of the $c\bar c$ pair. The predicted ratio of  branching ratios,  $BR(\psi' \to  \rho\pi)/BR(J/\psi \to \rho\pi) \sim 12\% $ appears, however, to be significantly above the 
     the measurements, which determine this ratio to be of the order of $1\%$. The 
      di-pion spectrum is dominated by the $\rho(770)$ resonance and  this so-called $\rho-\pi$ puzzle still remains largely unresolved~\cite{rp1,rp2,rp3,rp4,rp5}.  To better understand its origin may require gaining further information about $c\bar c$ wave functions and/or light quark production dynamics. These can be determined, at least indirectly,  by comparing microscopic model predictions with measurements of  charmonium couplings to light quark resonances other then the $\rho(770)$ meson. In this paper we discuss methods  for determining these couplings. 
     
 To be able to determine which resonances are produced in a given reaction it is necessary to perform a partial wave analysis  (PWA). While the full reaction amplitude is a function of the energies, momenta and helicities of external particles, partial waves emerge 
  after the amplitude is projected onto waves with well defined angular momenta. These are associated with resonances appearing  in the intermediate states. In data analysis,  partial waves are often used from the start without reference to the underlying reaction amplitude.  
   A finite sum of partial waves, however,  cannot reproduce   singularities of the full amplitude and  analyses based solely on a  model for partial waves are insensitive to a large set of dynamical constraints. Without  prior  knowledge, using only the energy dependence  of a single partial wave, it is difficult to determine  resonance parameters 
   unambiguously. On the other hand  the  full amplitude does in principle contain information about all resonances. 
     In  particular the asymptotic behavior in the cross-channel energy variable is given by the leading Regge pole of the direct channel  partial waves.  Partial waves possess a rich  analytical structure which 
       extends beyond the energy dependence at fixed angular momentum. They are analytical functions of complex 
        angular momentum and the motion of singularities in  the angular momentum plane as a function of 
       energy gives  a connection between resonances of varying spins and masses.  This connection is specific to the 
      underlying  dynamics responsible for resonance formation. For example, a linear rise of a Regge pole trajectory is a manifestation of confinement and is related  to the existence of an infinite number of bound states.

    While it is unknown how to construct reaction amplitudes that take full advantage of S-matrix constraints, when applicable,
     the Veneziano model and its various extensions are a good starting point in developing amplitude models. 
    In this paper, using the Veneziano model,  we  want to show how important it is to go beyond individual partial waves 
     and to illustrate the benefits of considering the full amplitude by applying the model to the specific case of charmonium decays. 

The paper is organized as follows. In Section  \ref{M} we give a brief description of the Veneziano model. There is an extensive literature on the subject and for more details we refer the reader to ~\cite{DS} and references therein. In Section ~\ref{rp} we discuss the procedure used to isolate selected poles. The amplitudes obtained in this section are used in Section ~\ref{an} to analyze di-pion mass distributions from charmonium decays. Summary and outlook are given in Section ~\ref{sum}.

\section{Generalized Veneziano Amplitudes} 
\label{M} 
Properties of the Veneziano model will be illustrated by considering the decays of vector charmonia, $J/\psi$ and $\psi'$ to three pions, $\pi^+\pi^-\pi^0$.   For simplicity, we neglect the pion mass since $M_\psi^2 >> m_\pi^2$:  in units of GeV, $M_\psi^2 = O(10)$ is the mass squared of the decaying particle.  The generalization to other reactions that involve four external particles is in principle straightforward and a number of 2-to-2 and 1-to-3 process have been considered this way  in the past. It is worth noting that the original paper by Veneziano~\cite{V} deals with another vector-to-three pion decay, namely $\omega \to 3\pi$. The main difference between $\omega$ and charmonium decays is that in the latter the $3\pi$ phase space is significantly larger  and direct production of several di-pion resonances above the $\rho(770)$ is possible.   

In the Veneziano model the amplitude $A$, describing a decay of a vector meson with momentum $p$ and helicity $\lambda$ to three pions,  $V(p,\lambda) \to \pi^i(p_1) \pi^j(p_2)\pi^k(p_3)$, after the isospin tensor $\epsilon_{ijk}$ has been factored out,  is given by 
\begin{equation}
A(s,t,u)  =  K  [ A_{n,m} (s,t) + A_{n,m}(s,u) + A_{n,m}(t,u)]. \label{A} 
\end{equation} 
 Here $s,t$ and $u$ are the standard Mandelstam variables,  $s + t + u = M_\psi^2$ and the scalar functions $A_{n,m}$ are given by 
   \begin{equation} 
A_{n,m}(s,t)  \equiv \frac{ \Gamma(n - \alpha_s)\Gamma(n-\alpha_t)}{\Gamma(n + m - \alpha_s -\alpha_t)},
\end{equation} 
 with $n,m$ being positive integers and  $1 \le m \le n$. 
  The lower limit on $m$ guarantees that $A(s,t,u)$ has the expected high-energy behavior (see below) and the upper limit eliminates double poles in overlapping channels. 
  The leading, linear Regge trajectory  $\alpha(s)  = \alpha_0 + \alpha' s$ is denoted by  $\alpha_s$  for 
  short. Finally  $K$ is a kinematical factor, 
\begin{equation} 
K =  \epsilon_{\mu\nu\alpha\beta} \epsilon_\mu(p,\lambda) p_1^\nu p_2^\alpha p_3^\beta
\end{equation}
 originating from presence of an odd-number of pions (unnatural parity) in the final state. 

\begin{figure}
\includegraphics[scale=0.24,angle=0]{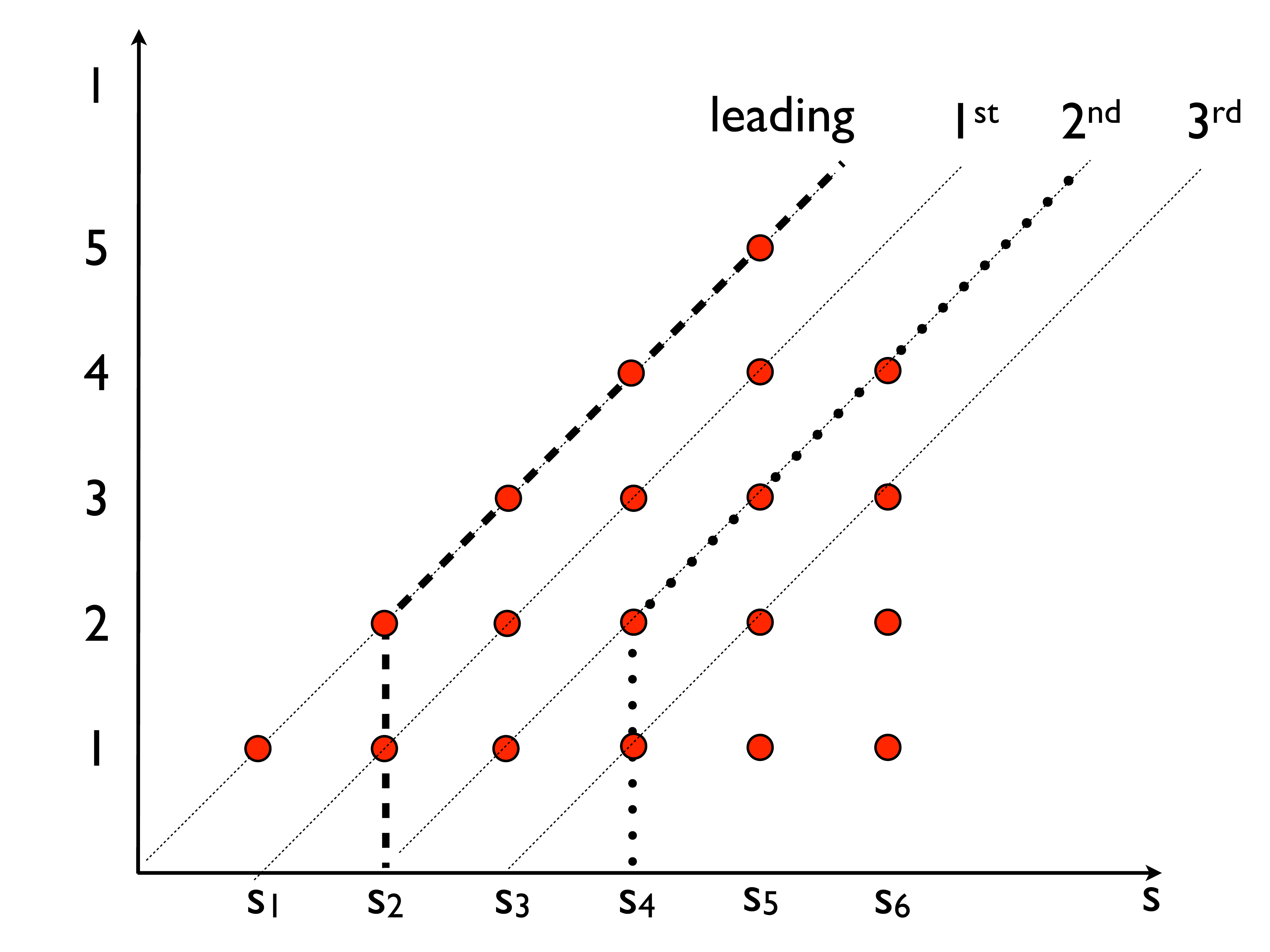}
 \caption{\rm \label{fig-spectrum} 
Spectrum in the $s$-channel of the generalized Veneziano amplitude model of Eq.~(\ref{A}). The leading and daughter Regge trajectories are marked by thin solid lines and the resonances  by dots at integer values of the spin $l$ marked on the vertical axis. 
The dashed and dotted lines are drawn to illustrate which  resonance contribute to two amplitudes, chosen to be  
$A_{2,1}$ and $A_{4,3}$, respectively. All  (infinite number) of resonances on, and to the right of the dashed line contribute to $A_{2,1}$, and all resonances on, and to the right of the dotted line contribute to $A_{4,3}$. In the plot, parameters of the trajectory, $\alpha_0$ and $\alpha'$ were chosen arbitrarily. } 
\end{figure}

The  Veneziano formula exhibits  the expected behavior of the amplitude in the large-$N_c$  limit, with the QCD boson  spectrum  saturated by narrow resonances and confinement resulting  in linear Regge trajectories. This  spectrum is manifested in the singularities of the amplitudes $A_{n,m}(s,t)$, which have 
 simple poles. For given $n$, there is an infinity of $s$-channel 
  poles labeled by a nonnegative integer $k$ that are located at $s= s_{n+k}$, satisfying 
 
\begin{equation} 
\alpha(s_{n+k}) = n + k. \label{pole} 
\end{equation} 
In the vicinity of the pole the amplitude $A_{n,m}(s,t)$ is given by 
\begin{equation} 
A(s \sim s_{n+k}) =  \frac{\beta_{n,m,k}(t)}{s_{n+k} - s} 
\end{equation} 
where  the residue,  
\begin{equation} 
 \beta_{n,m,k}(t) = \frac{(-1)^{k}}{\alpha' k!} \frac{\Gamma( n - \alpha_t) }{\Gamma(m - k  -\alpha_t)} \label{b}
 \end{equation} 
is a polynomial in $t$ of the order $L_{max}  \equiv k + n - m \ge 0$. 
We thus conclude that for each $k$ the full amplitude of Eq.~(\ref{A}), in each channel ($s$, $t$ and $u$), describes a finite 
number of degenerate, narrow (zero) width resonances that have spins  in the range,  
$ 1 \le l \le L_{max}+1$.   The  additional  unit of angular momentum arrises from the angular dependence of the kinematic factor $K$. 
       The integers $n$ and $m$ determine which resonances contribute (poles) to the amplitude. 
  It follows from  Eqs.~(\ref{pole},\ref{b}) that amplitudes with $m=1$, {\it i.e.} $A_{n,1}$, have poles whose location is determined by the leading trajectory and from  
 all  subsequent  daughter trajectories. The amplitudes  $A_{n,2}$ have poles originating from 
   the $1^{st}$ daughter and subsequent daughters, $A_{n,3}$ from the $2^{nd}$ and all  subsequent daughters, {\it etc.} 
 The daughter trajectories are defined by, 
   \begin{equation} 
   \alpha^{(m)}(s) \equiv \alpha(s) - (m-1). 
   \end{equation} 
 So that the leading trajectory  $\alpha(s)$ corresponds to $\alpha^{(1)}(s)$,  $\alpha^{(2)}$ is the $1^{st}$ daughter 
  and so on. The trajectories and the spectrum are illustrated in Fig.~\ref{fig-spectrum}. 
  For fixed-$t$, the asymptotic behavior of $A_{n,m}(s,t)$ at large-$s$ reflects the presence of an infinite number of 
   resonances in the $t$-channel.   Using Stirling's formula one finds, 
 \begin{equation} 
A_{n,m}(s \to \infty,t) \propto \frac{1}{s} \Gamma(n - \alpha_t) (-s)^{ \alpha^{(m)}_t}.   \label{as1} 
\end{equation} 
For large-$s$ the kinematical factor in Eq.~(\ref{A}) contributes an additional power of $s$ so that, the full amplitude has the expected Regge limit, 
\begin{equation} 
A(s,t,u) \propto (-s)^{\alpha(t)}  \label{regge} 
\end{equation} 
 that arises from the leading, $m=1$ trajectory. 
 
 \section{ Removal of poles} 
 \label{rp} 
As described in the preceding section, for given $n$ and $m$ the amplitude $A_{n,m}$ contains an infinite number of poles.  
   Since production of resonances is reaction dependent, it is necessary to find a mechanism  
 that allows for the residues to be process dependent and in particular for the possibility that some of them vanish if 
  a resonance formation is forbidden, {\it e.g} by  a conservation law.  One possibility is to consider linear combinations  
    \begin{equation} 
A_{n,m}(s,t) \to {\cal A}(s,t) = \sum_{n\ge 1,n \le m \le 1} c_{n,m} A_{n,m}(s,t). \label{c}
\end{equation} 
The coefficients $c_{n,m}$ need to be chosen in such a way that ${\cal A}$'s only couples to  resonances that contribute to the process in question. In the case considered here, of an isoscalar boson decaying to three pions, isospin conservation demands  each pair of  pions to be produced  in the isospin-1 state, which together with  Bose statistics forbids production of 
 spin-even resonances  in  $s$, $t$ and $u$ channels. 

One way to proceed  is  to  construct combinations of $A_{n,m}$'s that result in ${\cal A}$ containing only a finite number 
 of  Regge poles. As will be shown below, this requires an infinite number of terms in Eq.~(\ref{c}).   Alternatively one can attempt data analysis with a  finite number of  linear combinations of the $A_{n,m}$'s and let the fit determine the coefficients $c_{n,m}$~\cite{lovelace,alt,pok}. We find the first approach  more appealing particularly in the context of   resonance production.   Resonance properties are constrained by unitarity. This forces  Regge trajectories to be non-linear, but the Veneziano model relies on  trajectories  that are real and linear.   Even though there are extensions of the Veneziano model that introduce non-linear trajectories~\cite{Bugrij:1973ph,laszlo},  it is far simpler to implement unitarity  
at the level of an isolated Regge pole~\cite{ur1,ur2,Londergan:2013dza}.  An amplitude that contains only a finite number of resonance 
  poles, however,  does not reproduce the Regge limit in the crossed channel. And it is important to preserve the asymptotic behavior since it helps constraining the background  under directly produced resonances. Thus to take the full advantage of  the Veneziano model, we will, at the end,  need to consider an infinite number of poles.     

Before we impose the asymptotic behavior on the forms given by Eq.~(\ref{c}), it is nevertheless useful to ask what  choice of 
 $c_{n,m}$'s produces an ${\cal A}$ that contains only a finite number of resonance poles.   Since the $A_{n,n}$ amplitudes contain an infinite number of  poles, in order to cancel all, but a finite number of them,  an  infinite number of    $c_{n,m}$'s in  Eq.~(\ref{c}) must be 
 non-vanishing.
        It is not difficult to find a relation between the coefficients that decouples all, but a finite number of poles. Consider, for example,  keeping only the pole at $\alpha(s) = 1$, {\it i.e.} at $s = s_1$. 
   This pole is only  present in the amplitude $A_{1,1}$  since amplitudes with  $n>1$ have poles at $s_n \ge s_2$ ({\it cf.}  Fig.~\ref{fig-spectrum}).  
   There  is only one amplitude $A_{1,m}= A_{1,1}$ so a single coefficient $c_{1,1}$ determines the residue and ultimately 
    coupling to the pole at $s=s_1$.  
   The amplitude $A_{1,1}$, however, also has poles at higher masses located at  $\alpha_s = 2,3,\cdots$ with residues that are polynomials in $t$ of the order of $O(1),O(2),\cdots$, respectively. If we only want to keep the pole at $\alpha(s) = 1$, these higher mass poles of $A_{1,1}$ must be canceled by the same poles in amplitudes $A_{n,m}$ with $n \ge 2$. 
     Specifically, the pole in $A_{1,1}$ at $\alpha_s = 2$ can only by canceled by the same pole in the two amplitudes: $A_{2,m}$, $m=1,2$, since for $n>2$ no other $A_{n,m}$ contains this pole.  The amplitudes $A_{2,1}$ and $A_{2,2}$  are polynomials in $t$ of the order of  $O(1)$ and $O(0)$, respectively. We can therefore uniquely determine the  two coefficients,  $c_{2,1}$ and $c_{2,2}$ in terms of $c_{1,1}$  so that the first order polynomial in $t$ at the $s=s_2$ pole of $A_{1,1}$ is identical to the first order polynomial   
       $t$ at the pole of $A_{2,1}$ and $A_{2,2}$. This way we can make the sum of residues between the three amplitudes, $A_{1,1}, A_{2,1}$ and $A_{2,2}$ at the pole $\alpha(s) = 2$  vanish identically. 
          Similarly, at the $\alpha_s = 3$ pole of $A_{1,1}$, the residue is 
    an $O(2)$ polynomial in $t$. This pole is also present in $A_{2,1}$ and $A_{2,2}$ with residues order, $O(2)$ and $O(1)$ polynomials, respectively, and it is also present in $A_{3,1}$, $A_{3,2}$ and $A_{3,3}$ with residues of the order of $O(2)$, $O(1)$ and $O(0)$, respectively. With $c_{2,1}$ and $c_{2,2}$ already fixed, $c_{3,1}$, $c_{3,2}$ and $c_{3,3}$ are now uniquely determined in terms of  $c_{1,1}$ by the requirement that the total residue of the $\alpha_s = 3$ pole, which is an $O(2)$ polynomial in $t$, vanishes. Continuing in this way all poles in $s$  satisfying $\alpha(s) > 1$ can be eliminated. It is easy to check that this is achieved by setting, for $n \ge 2$ 
    \begin{equation} 
    c_{n,1} = \frac{c_{1,1}}{\Gamma(n)},\;  c_{n,2} = - \frac{c_{1,1}}{\Gamma(n-1)}, \;  c_{n,m} = 0  \mbox{ for }m \ge 3.  \label{c-1} 
    \end{equation} 
 The resulting amplitude ${\cal A}$ is then given by
   \begin{equation} 
   {\cal A}_1(s,t) = c_{1,1} \frac{2- \alpha_s -\alpha_t}{(1- \alpha_s)(1-\alpha_t)}. \label{a=1}
 \end{equation} 
 where the subscript indicates the  location of the pole. 
  This simple result could have been anticipated. The combination of the $\Gamma$ functions in $A_{n,m}(s,t)$ can be written as an infinite sums of simple poles in $s$. The amplitude is symmetric in $s$ and $t$, therefore if all poles but the one at $\alpha_s=1$ are left, by $s\leftrightarrow t$ symmetry ${\cal A}$ must also contain a pole in $t$ but not a double pole. This leaves Eq.~(\ref{a=1}) as the only possibility. 
 As expected once the infinite number of poles has been eliminated ${\cal A}$ no longer exhibits the Regge limit. 
   We will return to this point in the following subsection.

  This elimination procedure can be generalized to produce amplitudes with isolated 
   poles at any higher, integer value of $\alpha_s$ 
   For example, to construct an amplitude with a single pole in $s$ at $\alpha(s) = 3$, one starts with the three  amplitudes $A_{3,m}$, $m=1,2,3$ and determines the coefficients $c_{n,m}$ for $n \ge 4$ in terms of $c_{3,1}$, $c_{3,2}$ and $c_{3,3}$ that  remove all poles at $\alpha(s) > 3$.  The most general structure of the amplitude with the pole at $\alpha(s)=3$ only is therefore given by 
       \begin{equation} 
 {\cal A}_3(s,t) =   \frac{ (6 - \alpha_s - \alpha_t)}{ (3 -\alpha_s)(3 - \alpha_t)}
 \sum_{i=1}^3 a_{3,i} (-\alpha_s - \alpha_t)^{i-1}. 
  \label{a=3}  
 \end{equation}
 The first factor in the numerator guarantees that ${\cal A}_3$ does not have the double pole at $\alpha_s = \alpha_t = 3$. It is 
  followed by a product of two monomials  in $\alpha_s + \alpha_t$ that generate $O(2)$ polynomial in $s$ or $t$ at the pole 
   located at  $\alpha_t = 3$ or $\alpha_s =3$, respectively.  
 Having three  parameters $a_{3,m}$, $m=1,2,3$ determining the amplitude ${\cal A}_3(s,t)$ enables to decompose the residue in terms of  an arbitrary linear combination of  partial waves with $l=0,1,2$. We note, however, that once the kinematic factor $K$ 
   is taken into account, {\it cf.} Eq.~(\ref{A}),  the $\alpha(s) = 3$ pole actually represents  (narrow) resonances with spins 
    $l=1,2,3$. The coefficients $a_{3,m}$, $m=1,2,3$ can therefore be chosen to decouple the $l=2$ isobar. 
 In general we find that an amplitude  ${\cal A}_n$, which has a single pole at $\alpha_s = n$ or $\alpha_t = n$  is given by 
        \begin{equation} 
 {\cal A}_n(s,t) =   \frac{ (2n - \alpha_s - \alpha_t)}{ (n -\alpha_s)(n - \alpha_t)}
 \sum_{i=1}^n a_{n,i} (-\alpha_s - \alpha_t)^{i-1}. 
  \label{a=n}  
 \end{equation}

  \subsection{Regge assymptotics} 
The large-s behavior of the amplitude ${\cal A}_n$, is given by  $s^{n-1}$. 
    The expected, Regge asymptotic behavior,  however,  should be 
         $s^{\alpha(t)-1}$, {\it cf.} Eqs.~(\ref{as1},\ref{regge}). The Regge behavior can only emerge from an infinite number of poles, therefore we need to modify the procedure outlined above and allow for an infinite number of poles to be present in ${\cal A}$. 
          If we include an infinite number of poles located at say, $n > N$ where $N$ is chosen such that  $N >> \alpha' M^2_{\psi}$, ($\alpha' \sim 0.9\mbox{ GeV}^{-2}$ is the Regge trajectory slope), these poles will contribute a smooth background in the decay region.  
    
          With the  $c$'s given by Eq.~(\ref{c-1}) and the sum over $n$ in Eq.~(\ref{c}) truncated at $n=N$ we find that instead 
         of  a single pole at $\alpha=1$ we obtain 
          \begin{eqnarray} 
   {\cal A}_1(s,t)  & \to  &     {\cal A}_1(s,t;N)  =   a_{1,1}\frac{ 2 -\alpha_s - \alpha_t}{(1 - \alpha_s)(1 - \alpha_t)}  \nonumber \\
        &\times  & \frac{ \Gamma(N + 1 -\alpha_s)\Gamma(N+1-\alpha_t)}{\Gamma(N)\Gamma(N+2 - \alpha_s - \alpha_t)}. \label{with-r}  
          \end{eqnarray} 
         For $s >> N$ this amplitude has the desired Regge behavior $\propto s^{\alpha(t) - 1}$. As expected it is free from poles in the  range    $2 \le  \alpha(s)  \le  N$ and of course the same holds in the $t$-channel. For  $N$ large enough 
            {\it i.e.} $N >>  \alpha' M_\psi^2$ the contribution to the decay region, of the undesired, high-energy poles  located at $\alpha \ge N+1$      is power suppressed 
         \begin{eqnarray}
         {\cal A}_1(\{s,t\}<M_\psi^2;N) & = &  a_{1,1}\frac{ 2 -\alpha_s - \alpha_t}{(1 - \alpha_s)(1 - \alpha_t)} \nonumber \\
 &\times  &          \left[ 1 + O\left( \frac{\alpha' M_\psi^2}{N} \right) \right] 
         \end{eqnarray}
         thus as mentioned above, can be interpreted as background. 
The generalization of Eq.~(\ref{with-r}) to an amplitude, which in the decay region has a pole at $\alpha=n$, {\it i.e.} generalization of Eq.~(\ref{a=n}) to an amplitude with proper Regge asymptotics 
 is given by 

\begin{eqnarray}
& &  {\cal A}_n(s,t;N)  =    \frac{ 2n - \alpha_s - \alpha_t}{(n - \alpha_s)(n - \alpha_t)} 
   \sum_{i=1}^{n} a_{n,i} (-\alpha_s - \alpha_t)^{i-1} \nonumber \\
& & \times   \frac{ \Gamma(N + 1 -\alpha_s)\Gamma(N+1-\alpha_t)}{\Gamma(N + 1- n
)\Gamma(N+n+1 - \alpha_s - \alpha_t)}. \label{af}
\end{eqnarray}

In the following we use these amplitudes to describe  $J\psi$ and $\psi'$, three pion decays.          
         
         \section{ Application to vector charmonium decays} 
         \label{an} 
   
   Both, the $J/\psi$ and $\psi'$ decays show a clear signal of  $\rho(770)$ production. In additional there is indication 
   of higher mass resonance production in $\psi'$ decays.  This is not necessarily the case 
    in $J/\psi$ decays, nevertheless the single $\rho(770)$ does not saturate the spectrum either. In the past we attempted to describe  the $J/\psi$ decay distribution with additional partial waves.  We found  that interference effects are strong and even after adding       $\pi\pi$ interactions up to $\sim 1.6\mbox{ GeV}$ the description remained  quite poor.  Continuing to expand the partial wave basis to cover even higher mass region would lead to a quite unconstrained analysis.  
           On the other hand with the amplitudes developed in this paper,  all  partial waves are related to the same Regge trajectory and that gives a very strong constraint on amplitude analysis. 
              
        We will thus attempt to fit the di-pion mass distribution using a combination of amplitudes given by Eq.~(\ref{af}) truncated 
         up to some maximal value of $n=n_{max}$. For di-pion mass up to $\sim 3.5\mbox{ GeV}$ which is accessible in $\psi'$ decay, resonances with masses corresponding to $n$ up to $\sim 12$ can be directly produced. We have found however that when   using only the di-pion mass projection the fit  is quite insensitive to amplitudes with $n$ larger than $\sim 5-6$. In the following we will therefore truncate the sum over $n$ at $n_{max} = 6$. 
  As long as $N >>  \alpha' M_\psi^2$ we find little sensitivity to $N$, so w take  $N=20$, which is comfortably above the boundary of available phase space in both $J/\psi$ and $\psi'$ decays.  In terms of the $s$-channel partial waves, $f_l(s)$,  the scalar amplitude  $F \equiv  A/K$ in Eq.~(\ref{A}) 
\begin{equation} 
F(s,t,u) =  \sum_{n=1}^{n_{max}}\left[  {\cal A}_n(s,t;N) + {\cal A}_n(s,u;N) + {\cal A}_n(t,u;N) \right] \label{FF} 
\end{equation} 
is given by \cite{Niecknig:2012sj} 
\begin{equation} 
F(s,t,u) = \sum_{l} f_l(s) P'_l(z)  \label{pwa} 
\end{equation} 
where, ignoring the pion mass,  $z = (t-u)/(M_\psi^2-s)$ is the cosine of $s$-channel scattering angle and $P_l(z)$ are the Legendre polynomials. The $s$-channel pole of $A$ located at $\alpha(s) = n$ contains partial waves with $l=0,\cdots n$.  
 At the pole located at $\alpha(s) = n$  the partial waves $f_l(s)$ are given by $f_l(s) =  g_{n,l}/(n - \alpha(s))$  with 
\begin{equation} 
g_{n,l} = \int_{-1}^{1} \frac{dz}{2}[P_{l-1}(z) - P_{l+1}(z)] \mbox{Res} {\cal A}_n(n,t;N) + (t \to u) ] \label{g} 
\end{equation} 
where 
\begin{equation} 
\mbox{ Res} {\cal A}_n(n,t;N) =   \sum_{i=1}^{n} a_{n,i} (-n - \alpha_t)^{i-1} \equiv \sum_{i=1}^n a'_{n,i} t^{i-1}. 
\end{equation}
The residue $g_{n,l}$ is the product of two couplings. One of them is the coupling of the charmonium to  a di-pion resonance of spin $l$ and mass $m_r$ given be  
 $\alpha(m_r^2) = n$ and the other is the coupling of this resonance to the di-pion decay channel. 
 
Decoupling of the spin-even resonances implies that  $\mbox{ Res} {\cal A}_n(n,t;N)$ should be an even function of $t$. 
 For $n=1$ $\mbox{ Res} {\cal A}_n(n,t;N)  = a'_{1,1}$ and from Eq~\ref{g} we can determine charmonium coupling to the 
 $\rho(770)\pi$ intermediate state. For $n=1$,  we need to set $a'_{2,2}=0$ to decouple the $l=2$ wave. The $\alpha(m^2_r) = 2$ pole would then correspond to the first excitation of the  $\rho$-meson, {\it i.e.} the $\rho(1450)$. 
 For $n=3$ we set $a'_{3,2} = 0$ and  the pole at $\alpha(m_r^2) = 3$ describes the $l=1$, second excitation of the $\rho$, 
   {\it i.e.} the $\rho(1570)$ and the $l=3$,  $\rho_3(1690)$ resonance. 
   For $n=4$, $a'_{4,2} = a'_{4,4} = 0$ and we find two degenerate 
  resonances with masses given by $\alpha(m_r^2) = 4$, and spins $l=1,3$ that may be associated with $\rho(1900)$ and $\rho_3(1990)$, respectively.  Similarly the $\alpha(m_r^2) = 5$ pole, with $a'_{5,2} = a'_{5,4} = 0$ describes resonances with $l=1,3,5$, which can  correspond to $\rho(2150)$, $\rho_3(2250)$ and $\rho_5(2350)$, respectively. No higher mass $\rho$'s are known~\cite{pdg}. The pole at  $\alpha(m_r^2)=6$  produces  additional three  resonances with spins $l=1,3,5$ 
  and if the fit was robust, it would constitute a discovery of these states.

      \begin{figure}
\includegraphics[scale=0.5,angle=0]{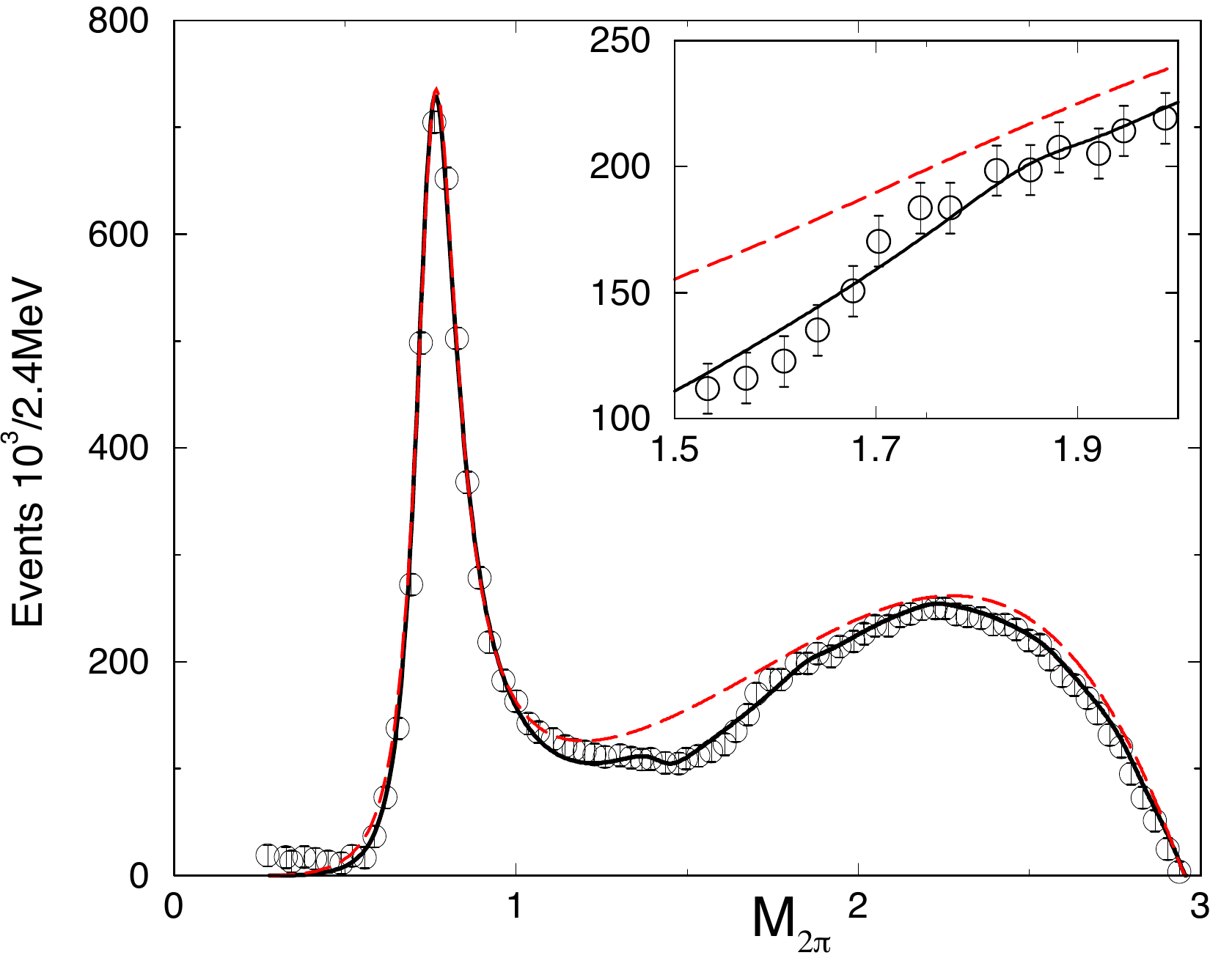}
 \caption{\rm 
Dalitz plot projection of the di-pion mass distribution from $J/\psi$ decay~\cite{8}. The solid is the result of the  fit with all, $n=1\cdots 6$ 
 amplitudes and the dashed line with the  ${\cal A}_1$ amplitude alone. The insert illustrates sensitivity to higher mass resonates in the $\alpha(m_r^2) = 3$ mass region. 
 \label{jpsi} }
\end{figure}

      \begin{figure}
\includegraphics[scale=0.5,angle=0]{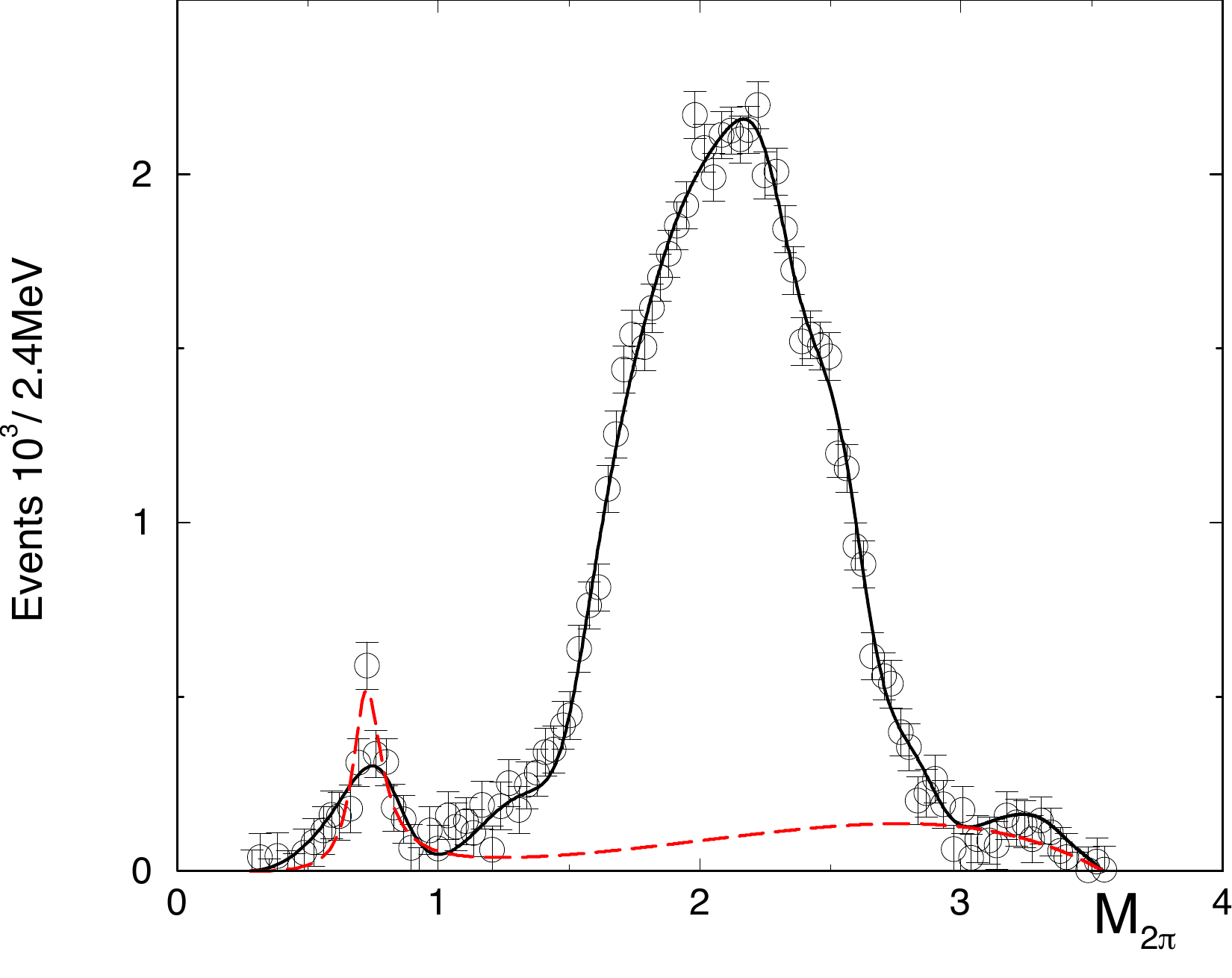}
\caption{\rm 
Dalitz plot projection of the di-pion mass distribution from $\psi'$ decay~\cite{8}. The solid is the result of the fit with all twelve amplitudes 
  and the dashed line is the fit with ${\cal A}_1$ alone. 
\label{psip} }
\end{figure}
 For each, channel, $J/\psi$ and $\psi'$ we thus fit twelve real parameters, $a'_{n,i}$ for $n=1\cdots 6$ and $i=1,\cdots n$, $i=even$ which via Eq.~(\ref{g}) determine production times decay coupling of the twelve di-pion resonances discussed above. 
In addition we allow the trajectories to be imaginary when appearing in the denominators of the ${\cal A}_n$'s in order  to be able to account for the finite  width of the resonances as required by unitarity.  
The $\rho$ trajectory is expected to be approximately equal to 
\begin{eqnarray} 
\alpha(s) & = &  1 +  \alpha'  (s - m_\rho^2) + i  \alpha'  m_\rho \Gamma_\rho \nonumber \\
 & \sim  &  0.47 + 0.9 s + 0.1 i\sqrt{s  -0.07}  \label{ar} 
\end{eqnarray}
where we also included the phase space factor, $\sqrt{s - 4m_\pi^2}$ in the imaginary part. 

The data and results of the fit are shown in Figs.~\ref{jpsi},\ref{psip}. The data is taken from for the resent measurement by the BESIII collaboration~\cite{8}. Unfortunately having no access to the Dalitz plot distribution we were only able to analyze the di-pion mass projection. Fitting mass projection carries larger systematic uncertainty compared to an event-by-event fit.   We therefore also  allow for the parameters of the trajectory, the intercept, slope and the magnitude of the imaginary part to float. The trajectory parameters, 

\begin{eqnarray}
\alpha(s)  & = &  (0.61 \pm 0.04) + (0.68 \pm 0.08)s \nonumber \\
 & + &  (0.11 \pm 0.02) i \sqrt{s - 0.07}
\end{eqnarray} 
obtained from the fit to the $J/\psi$ mass distribution are  in good agreement with Eq.~(\ref{ar}). As an estimate for systematic uncertainty we fit the $\rho(770)$ mass region with the $n=1$ amplitude alone. In this case we find 
\begin{eqnarray}
\alpha(s) &= &  (0.57 \pm 0.08) + (0.73 \pm 0.12)s \nonumber \\
& + &  (0.12 \pm 0.03) i \sqrt{s - 0.07}.
\end{eqnarray} 
The $J/\psi$ mass distribution is dominated by the $\rho(770)$ and by fitting the mass projection, in general we find weak sensitivity to the higher mass resonances. This is  reflected in large  uncertainties we obtain for the fit parameters $a'_{n,i}$ and 
 for this reason we do not attempt to determine resonance couplings. Nevertheless, examining Fig.~\ref{jpsi} it is clear 
 that the $\rho(770)$ meson alone is not capable of reproducing the data. Fitting data on an event-by-event basis 
   might be possible to obtain a more reliable estimate of higher mass resonance production. 
 
 

The  $\rho(770)$ is much less pronounced in the decay of the $\psi'$. If we try to determine $\rho(770)$ production alone by restricting the fit to the $n=1$ amplitude in the $M_{3\pi} < 1\mbox{ GeV}$ mass region we find 
\begin{eqnarray}
\alpha(s)  & = &  (0.45 \pm 0.51) + (1.0 \pm 0.9)s \nonumber \\
 & + & + (0.08 \pm 0.02) i \sqrt{s - 0.07}
\end{eqnarray} 
which is consistent with $J/\psi$ fit results but carries large statistical uncertainty. 
The fit with all six amplitudes gives 
\begin{eqnarray}
\alpha(s)  & = &  (0.55 \pm 0.2) + (0.65 \pm 0.1)s \nonumber \\
 & + & + (0.26 \pm 0.01) i \sqrt{s - 0.07}
\end{eqnarray} 
which well reproduces the real part but leads to a $\rho(770)$  width that is twice as large as observed. 
This is clearly seen in Fig.~\ref{psip}.  We find that the $\psi'$ seems to be dominated by resonances in the $\alpha = 5$ mass region, {\it i.e.} $\rho(2150)$, $\rho_3(2250)$ and $\rho_5(2350)$ seen as the large bump in Fig.~\ref{psip}. 
We expect that an event-by-event, likelihood fit would  remove the  large uncertainties we find in the couplings of these resonances~\cite{ryan}. 

 \section{\label{sum} Summary and Outlook} 

Based on the Veneziano model we constructed a set of amplitudes which {\it i)} isolate contributions from individual Regge trajectories, including the daughters, {\it ii)} preserve the asymptotic behavior emerging from  cross-channel Reggeons. 
The first property enables introduction of finite resonance widths, while the second helps avoid uncertainties in background parametrization common to analyses based on truncated partial wave expansion.   Given the limited sensitivity of a fit to  mass projections of the Dalitz plot, we retained the simple, linear parametrization of the real part of the trajectory, even though it introduces a high level of degeneracy between resonances. The partial wave amplitudes incorporate Regge poles. 
 This eliminates a freedom in choosing 
  which resonances to include in the analysis and gives a link between the measured spectrum and the underlying dynamics, {\it i.e.} in the large-$N_c$ limit the QCD spectrum is expected to match that of the Veneziano model. 
For precision studies based on event-by-event analysis, the approximations in  Regge trajectories can be easily eliminated  and  tailored to reproduce the data. Since Regge poles factorize, self-consistency can be tested by comparing resonance couplings between various decay modes containing the same set of resonance, {\it e.g.} $K\bar K\pi^0$. Extensions of the Veneziano approach beyond four-particles are known~\cite{multi}  and the approach discussed here can therefore be generalized to higher multiplicities.

\section*{Acknowledgments}
This work was supported in part by the U.S.\ Department of Energy under Grant No.~DE-FG0287ER40365.  The work was authored in part by Jefferson Science Associates, LLC under U.S. DOE Contract No. DE-AC05-06OR23177.

\end{document}